\documentclass[%
 reprint,
 amsmath,amssymb,
 aps,
]{revtex4-1}

\usepackage{graphicx}
\usepackage{dcolumn}
\usepackage{bm}
\usepackage{siunitx}

\begin{document}

\preprint{APS/123-QED}

\title{Stable liquid jets bouncing off soft gels}

\author{Dan Daniel$^{1}$}
\author{Xi Yao$^{2}$}
\author{Joanna Aizenberg$^{1,3,4}$}%
 \email{jaiz@seas.harvard.edu}

\affiliation{$^{1}$ John A. Paulson School of Engineering and Applied Sciences, Harvard University, Cambridge, MA 02138}
\affiliation{$^{2}$ City University of Hong Kong, Tat Chee Avenue, Hong Kong}%
\affiliation{$^{3}$Wyss Institute for Biologically Inspired Engineering, Harvard University, 
    Cambridge, MA 02138, USA} 
\affiliation{$^{4}$Department of Chemistry and Chemical Biology, Harvard University, 
    Cambridge, MA 02138, USA} 

\begin{abstract}
A liquid jet can stably bounce off a sufficiently soft gel, by following the contour of the dimple created upon impact. This new phenomenon is insensitive to the wetting properties of the gels and was observed for different liquids over a wide range of surface tensions, $\gamma$ = 24--72 mN/m. In contrast, other jet rebound phenomena are typically sensitive to $\gamma$: jet rebounds off a hard solid (e.g. superhydrophobic surface) or another liquid are possible only for high and low $\gamma$ liquids, respectively. This is because an air layer must be stabilized between the two interfaces. For a soft gel, no air layer is necessary and jet rebound remains stable even when there is direct liquid-gel contact. 
\end{abstract}

\maketitle

The ability of surfaces to repel liquids, either in the form of droplets or jets, is of broad interests and has numerous applications \cite{quere2005non, wong2013interfacial, feng2006design}. The rebound of single impinging water droplets off a rigid solid substrate, be it a superhydrophic surface \cite{richard2002surface,bird2013reducing, liu2014pancake} or a leidenfrost solid such as dry ice \cite{antonini2013water} is a well-known phenomenon; the equivalent rebound of liquid jet, on the other hand, is relatively less-studied, despite the pervasiveness of liquid jets in various applications \cite{eggers2008physics}. It was shown, only in recent years, that a water jet can stably bounce off a superhydrophobic surface with minimal energy loss at a low Weber number, We $= \rho R U^2/\eta< 10$, where $\rho$, $R$, $U$, and $\eta$ are the density, radius, velocity, and viscosity of the jet, respectively \cite{celestini2010water}. However, even a slight decrease of the surface tension of the liquid destabilizes the jet, and no stable jets of low-surface-tension liquids were observed on superhydrophobic surfaces. The behavior of a soft material or a fluid in contact with a liquid droplet has also been widely studied, but its interaction with liquid jet is rarely so \cite{eggers2008physics, extrand1996contact, roman2010elasto, style2013patterning}. Stable jets have also been observed in the case of a liquid bouncing off a bath of the same liquid, but such rebound is only possible for low-surface-tension liquids \cite{collyer1976kaye, versluis2006leaping, lee2013leaping, thrasher2007bouncing}.

\begin{figure}[h!]
\includegraphics[scale=1.0]{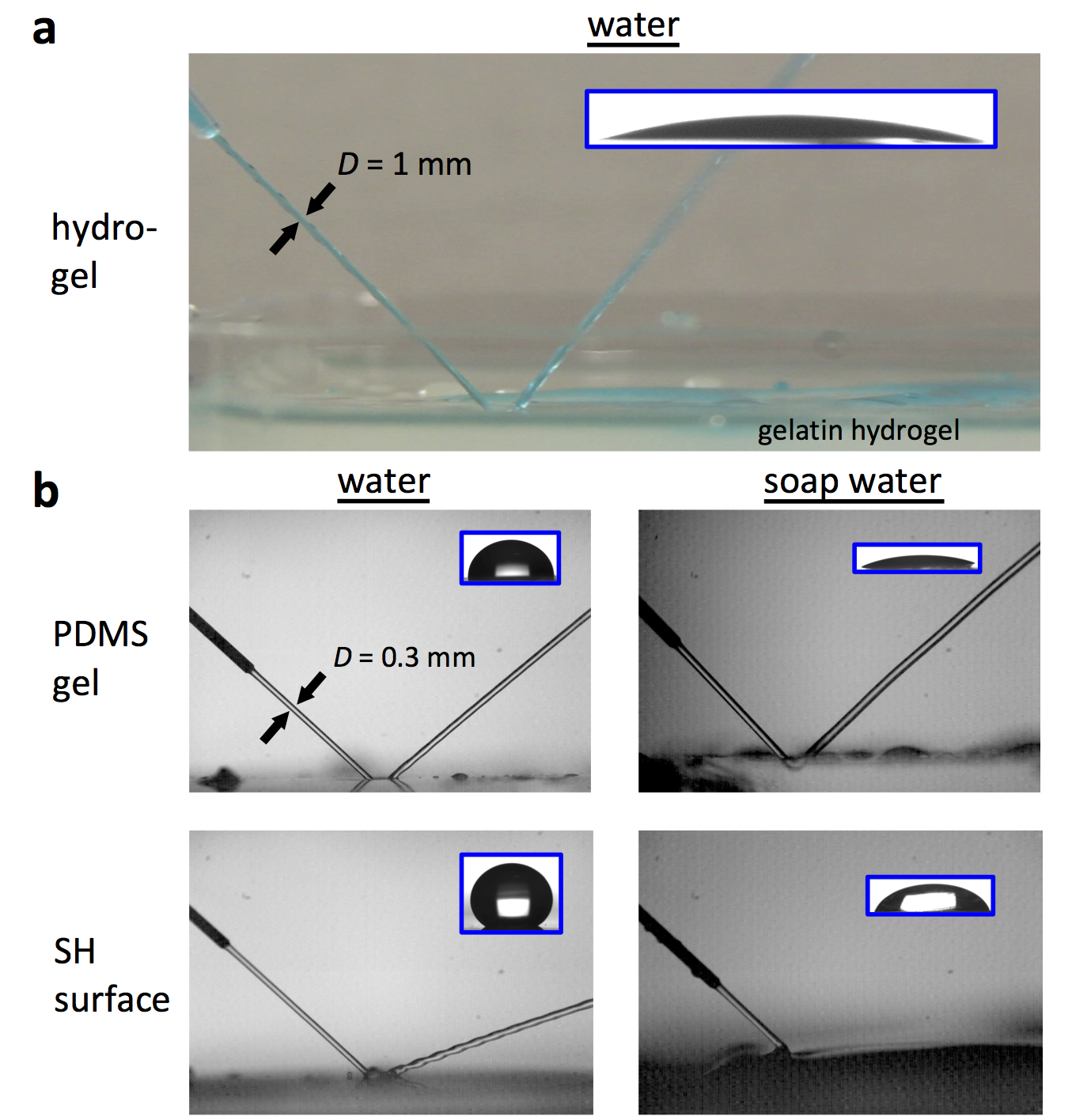}
\caption{\label{fig:gel} (a) A water jet can bounce off a hydrophilic gelatin gel that is 98.5 wt\% water. The inset shows the resulting contact angle of 16$^{\circ}$ for a \SI{10}{\micro\liter} water droplet. (b) A soft PDMS gel ($E$ = 1.2 kPa) can repel both water and soap-water, while a superhydrophobic (SH) surface loses its liquid repellence for soap-water. The corresponding contact angles of \SI{10}{\micro\liter} water/soap-water droplets shown in the insets are (left to right, top to bottom): $90^{\circ}$, $30^{\circ}$, $150^{\circ}$, and $60^{\circ}$. The SH surface was a hexagonal array of micropores monolayer of size $\sim $ \SI{1}{\micro\meter}, with static contact angle for water $\theta = 150^{\circ}$, and contact angle hysteresis $\Delta\theta = 10^{\circ}$. Detailed methods to fabricate the SH surface and PDMS gels used in this study are described elsewhere \cite{vogel2013transparency, Phil2014Gel}. 
}
\end{figure}


All the cases described above share one common feature: the rebound of droplets or jets is facilitated by a cushion of air layer between the two interfaces (liquid-solid or liquid-liquid). When this air layer becomes unstable or disrupted, for example due to increased wettability of the solid or the presence of dirt/defects along the interface, the rebound phenomenon is suppressed \cite{yarin2006drop, deng2013liquid, kolinski2014drops}. In this letter, we describe the mechanism and stability of jet rebound as it impacts a soft gel. This stable rebound phenomenon was observed for liquid jets over a wide range of surface tensions, $\gamma$ = 24--72 mN/m, and for a variety of soft gels.  We show that this phenomenon does not rely on the formation of the air cushion to stabilize the gel or the wetting characteristics of the liquids; it is driven by the deformability of the soft substrate, and therefore applicable to a wide range of soft, elastomeric materials, irrespective of their surface energies.

We begin the demonstration of this new phenomenon by showing that a water jet (diameter $D \sim$ 1.0 mm, velocity $U \sim$ 1 m/s) can bounce off a gelatin gel that is 98.5 wt\% water (Fig.~\ref{fig:gel}(a), see also Supplementary Movie S1); in other words, a stable high-surface-tension liquid jet bouncing off a high-surface-energy substrate containing the same liquid. The inset of Fig.~\ref{fig:gel}(a) shows a \SI{10}{\micro\liter} water droplet sitting on the gel with a contact angle of 16$^{\circ}$, i.e. the gel is hydrophilic. The same rebound phenomenon was observed not only for other hydrophilic hydrogels, such as polyacrylamide, but also for hydrophobic gels, such as polydimethylsiloxane gels (PDMS, 78--82 wt\% silicone oil, contact angle of water = $90\pm 5^{\circ}$). Unlike superhydrophobic (SH) surfaces that lose their repellence for low $\gamma$ liquid, the PDMS gel was able to repel liquid jet with $\gamma$ as low as 24 mN/m. Fig.~\ref{fig:gel}(b) illustrates this: soap-water jet ($D$ = 0.3 mm, $U \sim 2$ m/s, $\gamma$ = 30 mN/m) could bounce off the PDMS gel (78 wt\% silicone oil), but not the SH surface. The water jet, in comparison, was repelled by both surfaces. The insets show \SI{10}{\micro\liter} droplets of either water or soap-water sitting on the corresponding surfaces. 

To uncover the mechanism of such a stable jetting phenomenon and to avoid the complications of a hydrogel expanding/contracting due to absorption/evaporation of water, we chose to concentrate on jet rebound behavior on PDMS gels. Silicone oil is immiscible with water/ethanol and has a low vapor pressure $\sim$ 5 mm Hg. The Young's modulus $E$ of a PDMS gel can also be varied easily by using different wt\% of silicone oil. In our experiment, we used 78, 80, and 82 wt\% silicone oil gel to obtain $E$ = 0.6$\pm$0.1, 1.5$\pm$0.3, and 5$\pm$1 kPa, respectively. The ratio (mass) of sylgard 184 PDMS base and curing agent was kept at 1:1. For data presented below, the value of $E$ for each gel sample was measured individually using a rheometer. The PDMS gels used here behave mostly as elastic solids; the viscous component can be safely ignored, since $G''/G' \sim 0.05$, where $G''$ and $G'$ are the loss and elastic moduli.  

\begin{figure}[h!]
\includegraphics[scale=1.0]{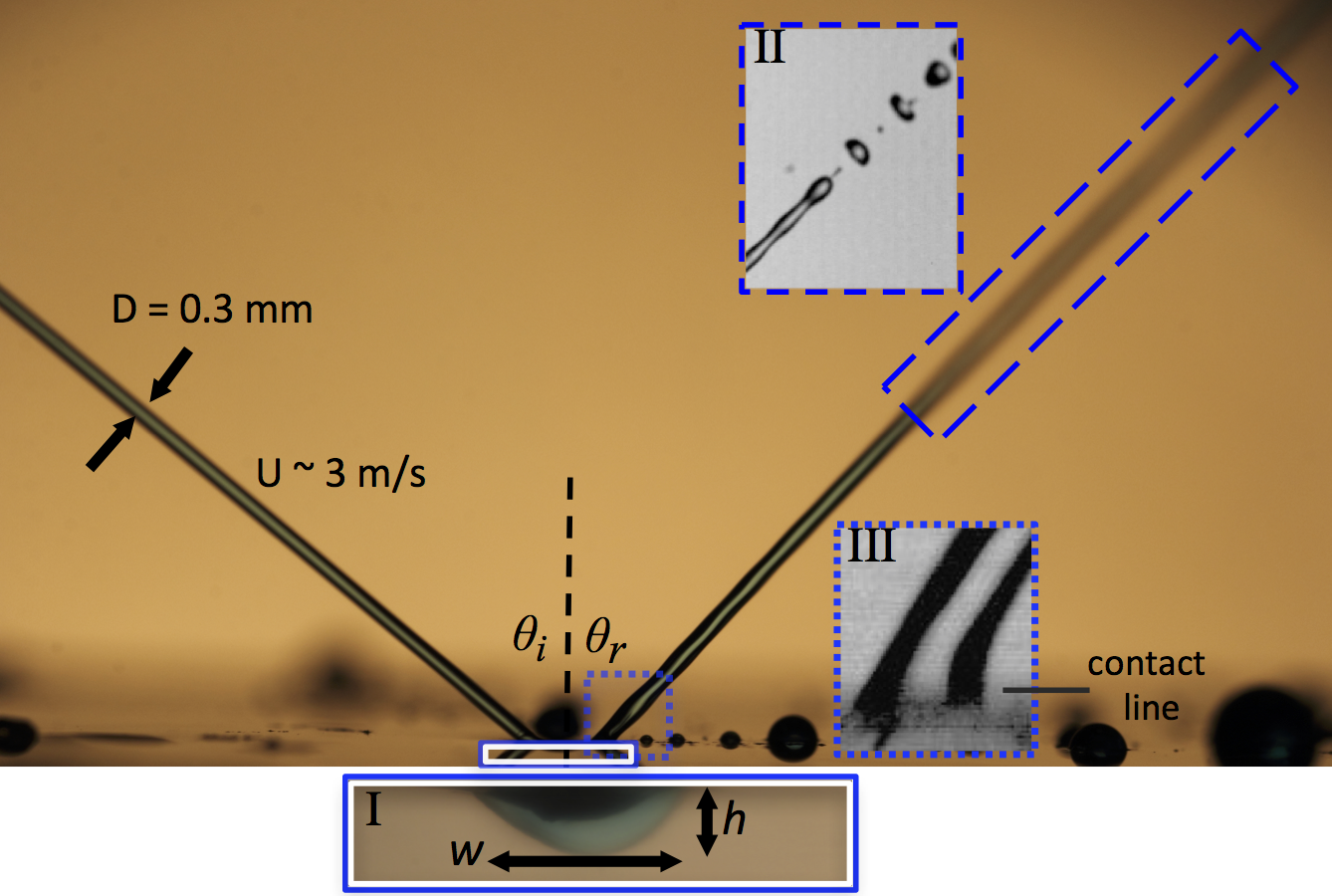}
\caption{\label{fig:schematic} The pressure of impacting jet creates a dimple of depth $h$ and width $w$ on the PDMS gel (Inset I), facilitating jet rebound. Inset II shows the eventual break-up of jet into droplets (Plateau-Rayleigh instability), while inset III shows the three-phase contact line of an exiting EtOH jet. Note that inset I was taken at a different focal plane from the jet above it, because the refractive index of the PDMS gel/silicone oil introduced an additional optical path.}
\end{figure}

\begin{figure}[h!]
\includegraphics[scale=0.9]{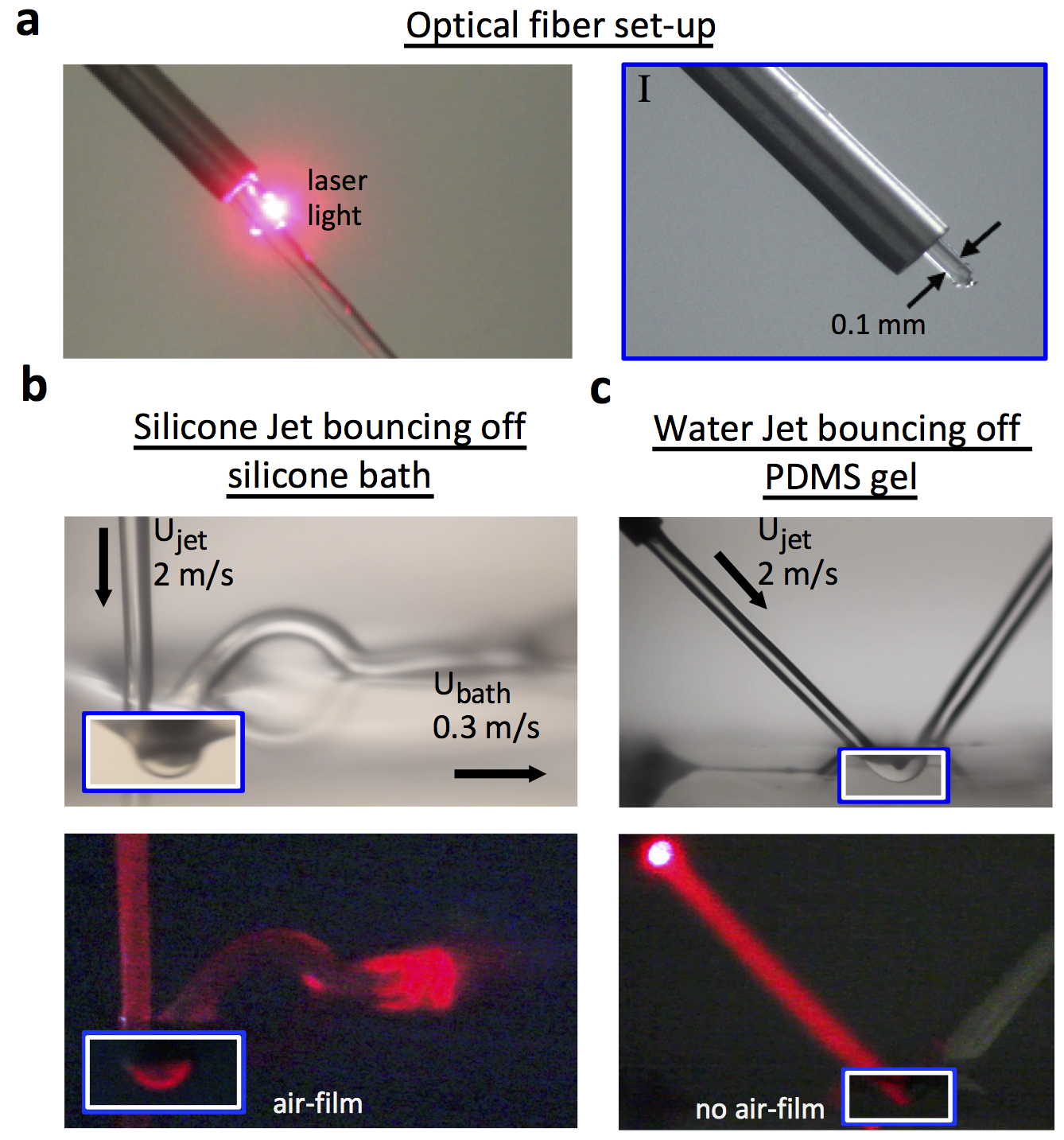}
\caption{\label{fig:laser} (a) An optical fibre with clad diameter of 0.1 mm (Inset I) was placed coaxial with the needle to couple laser light into the liquid jet. (b) Laser light (red) followed the contour of the silicone oil jet as it bounced off a silicone oil bath, because of a stable air layer that separated the two. (c) In contrast, there was no air layer separating water jet bouncing off a PDMS gel. Hence, the laser light passed through undeflected into the PDMS gel. N.B. The dimple created upon impact (boxed in (b) and (c)) was taken at a different focal plane from the jet above it, because the refractive index of the PDMS gel/silicone oil introduced an additional optical path.} 
\end{figure}

When a jet impacts a PDMS gel obliquely at an angle $\theta_{i}$ and speed $U$, a dimple of depth $h$ and width $w$ is formed on the gel surface (inset I, Fig.~\ref{fig:schematic}). The jet then follows the contour of the dimple and bounces off at an angle $\theta_{r}$, while retaining its cylindrical shape. After travelling a certain distance, the liquid jet then breaks up into droplets of size $\sim D$ and satellite droplets of size $\ll D$, due to Plateau-Rayleigh instability (inset II, Fig.~\ref{fig:schematic}; see also Supplementary Movie S2). We note that while a similar mechanism---dimple formation, followed by rebound---has been observed in the case of a liquid jet (Newtonian and non-Newtonian) bouncing off a bath of the same liquid, such rebound was only possible for low $\gamma$ liquids, where there is a stable air layer separating the liquid jet and bath \cite{collyer1976kaye, versluis2006leaping, lee2013leaping, thrasher2007bouncing}. The existence of this air layer is crucial for liquid-liquid rebound; without it, the liquid jet (e.g. water) simply merges with the liquid bath \cite{thrasher2007bouncing, lorenceau2004air, wadhwa2013noncoalescence}.

\begin{figure*}
\includegraphics[scale=0.9]{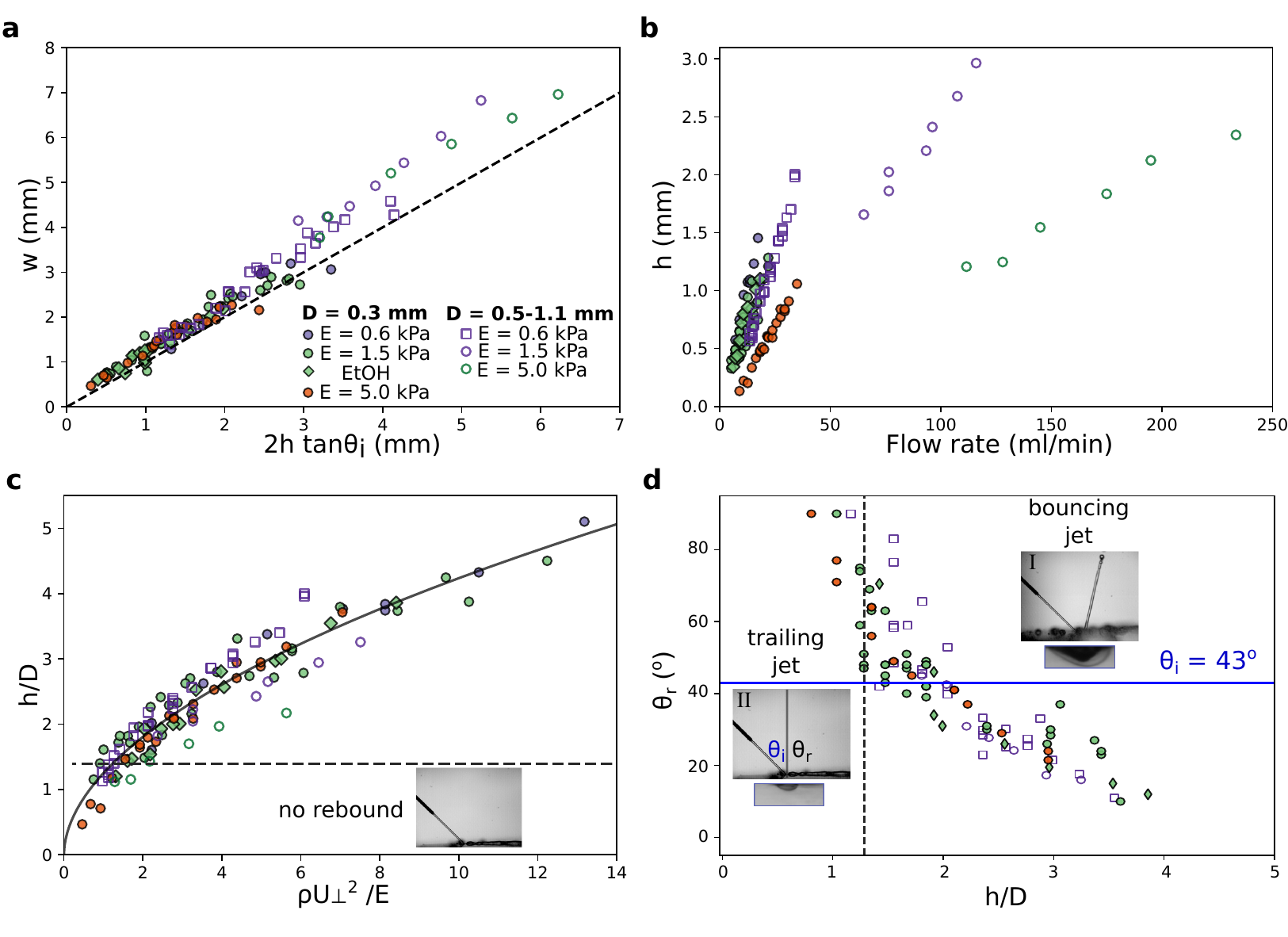}
\caption{\label{fig:deformation} (a) Dimple width $w$ plotted against $2h\tan{\theta_{i}}$, where $h$ is the dimple depth and $\theta_{i}$ is the incident angle for liquid jets (diameter $D$ = 0.29--1.1 mm) impacting PDMS gels (young's moduli $E$ = 0.6--1.5 kPa) at flow rates $Q$ = 5-250 ml/min. The liquid used was water, except for the data-points labeled with green diamonds, which were obtained for 30 wt\% ethanol (surface tension $\gamma$ = 35 mN/m). (b) $h$ increases with increasing flow rate $Q$, albeit at different rates depending on exact experimental conditions. (c) The results in (b) can be collapsed into one curve, when $h/D$ is plotted against $\rho U_{\perp}^2/E$, where $\rho$ is the density of the liquid and $U_{\perp}$ is the perpendicular component of the jet speed, i.e. $U\cos{\theta_{i}}$. (d) The rebound angle $\theta_{r}$ for a fixed $\theta_{i}$ is a function of the ratio $h/D$. The transition from a bouncing jet (inset I) to a trailing jet (inset II) occured at around $h/D$ = 1.4.}  
\end{figure*}

In comparison, jet rebound off a soft gel does not require an air layer; experimentally, no air layer was detected. For low $\gamma$ liquids, such as ethanol solution (70 wt\%, $\gamma$ = 24 mN/m), a three-phase contact line was always observed where the jet exited the dimple, ruling out existance of a continuous air layer (Inset III, Fig.~\ref{fig:schematic}). For a water jet, this three-phase contact line was not readily observable. However, we were able to confirm the absence of the air layer by shining laser light into the the water jet (Fig.~\ref{fig:laser}(a)). If there were an air layer---as is the case for a liquid jet (e.g. silicone oil) bouncing off a liquid bath---the liquid jet would act as an optical fiber and the laser light would remain inside the jet, following its contour, due to total internal reflection (Fig.~\ref{fig:laser}(b)). In contrast, for a water jet bouncing off a PDMS gel, the laser light passed through undeflected into the PDMS gel (Fig.~\ref{fig:laser}(c)). This suggests that either there was no air layer or that the air layer was nanometric in size, such that there was evanescent wave coupling between the liquid jet and the gel. The latter seems implausible since at such a thickness, the air layer can easily be destabilized by van der Waals' interactions. A small amount of alumina particles and milk (0.05 wt\% and 0.01 wt\%, respectively) was added to silicone oil and water jets to act as light scatterers so that the laser path inside the jet could be visualized, with insignificant change to the surface tension (measured $\Delta \gamma <$ 0.2 mN/m). 

Intuitively, we should expect the rebound behavior of jet to depend on the size and geometry of the dimple. For example, just to accommodate the jet, there should be a threshold dimple size $h_{\textrm{min}}, w_{\textrm{min}} \sim D$, for jetting to occur. For small dimple, $(w/2)/h \approx \tan\theta_{i}$, i.e. $w \approx 2h \tan\theta_{i}$. This is well-obeyed when $w < 2$ mm for a wide range of experimental conditons (Fig.~\ref{fig:deformation}(a)). Since this indentation is an elastic response to the pressure of water impact $\sim$ $\rho U^{2}$, we expect $h/D$ to depend on $\rho U_{\perp}^2/E$, i.e. $h/D \sim (\rho U_{\perp}^2/E)^{\nu}$, where $\rho$ is the density of the liquid, $U_{\perp} = U\cos{\theta_{i}}$ is the perpendicular component of the jet speed, and $\nu$ is an unknown scaling exponent. Note that according to the Buckingham-Pi theorem, there can only be two non-dimensionless groups, which we have chosen here to be $h/D$ and $\rho U_{\perp}^2/E$.  

The raw data (Fig.~\ref{fig:deformation}(b)) shows that $h$ is linearly proportional to flow rate $Q$, which suggests that the correct scaling should be $h/D \sim (\rho U_{\perp}^2/E)^{1/2}$, i.e. $\nu = 1/2$. This was verified experimentally for liquid jets of different surface tensions (water and 30 wt\% ethanol solution, $\gamma$ = 72 and 35 mN/m, respectively), different jet diameters $D$ = 0.29--1.1 mm, flow rates $Q$ = 5--250 ml/min, and impinging angles $\theta_{i}$ = 30--50$^{\circ}$ bouncing off PDMS gels with a range of Young's moduli $E$ = 0.6--5 kPa (Figs.~\ref{fig:deformation}(b), (c)). The best-fit line on Fig.~\ref{fig:deformation}(c) is the relation 
\begin{equation} \label{eq:def}
h/D = 1.25 (\rho U_{\perp}^2/E)^{1/2}.
\end{equation}
Since the elastic energy stored in the deformation is equal to the work done by the impacting jet, we expect $E \epsilon^{2} \sim \rho U_{\perp}^2$, i.e. $\epsilon \sim (\rho U_{\perp}^2/E)^{1/2}$, where $\epsilon$ is the characteristic strain in the system. Comparing this to equation (\ref{eq:def}), we find that $h/D \approx \epsilon$. Finally, $h_{\textrm{min}}/D$ was found to be 1.4, below which jet rebound becomes unstable (dashed black line, Figs.~\ref{fig:deformation}(c), (d)). 

There is some departure from the general trend of equation (\ref{eq:def}) for $h >$ 2.0 mm (see purple and green unfilled circles in Figs.~\ref{fig:deformation}(b), (c)), because $h$ starts to approach the thickness of the PDMS gel, which was kept at 8 mm for all the experiments, and the effects of the underlying hard, solid substrate become apparent. For a given incident $\theta_{i}$, the rebound angle $\theta_{r}$ is simply a function of $h/D$ (Fig.~\ref{fig:deformation}(d)), and below $h/D = 1.4$ (dashed vertical line), $\theta_{r}$ approaches $90^{\circ}$ precipitiously, i.e. jet transitions from bouncing (inset I) to trailing (inset II). Upon landing on the gel, the trailing jet initially moves in a straight line before starting to meander \cite{le2006meandering}.    

Looking at the data for 30 wt\% ethanol solution (green diamonds), we see that $\gamma$ does not affect dimple size/shape and rebound direction. This is because $h$ and $w$ $\sim$ mm; the effect of $\gamma$ becomes important only when the dimple size approaches the elasto-capillary length, $|S|/E$, where $|S|$ is the spreading parameter \cite{de2008capillarity, roman2010elasto}. Typically, $|S| \sim $ 10 mN/m, and for $E \sim$ 1 kPa, this length $\sim$ 10 $\mu$m \cite{duez2010wetting}.          

In conclusion, we have shown that a liquid jet with a wide range of $\gamma$ = 24--72 mN/m can bounce off a soft gel, irrespective of the gel's wetting properties and composition, by following the contour of the dimple formed upon impact. Our detailed experimental data and theoretical analysis show that the size and geometry of the dimple determine 1) whether jetting is possible and 2) the direction of jetting. We have further shown that this jet rebound phenomenon is possible even when there is direct liquid-gel contact, which is unlike other jet rebound phenomena that require a stable air layer separating the jet and the substrate. In this work, we have demonstrated how liquid-solid interactions can be controlled by changing the Young's modulus; this can have wide-ranging implications for the development of liquid-repellent surfaces and the science of wetting.    

We would like to thank Prof. Michael Brenner, Prof. L. Mahadevan and Prof. Kyoo-Chul Park for useful discussions. The work was supported partially by the ONR MURI Award No. N00014-12-1-0875 and by the Advanced Research Projects Agency-Energy (ARPA-E), U.S. Department of Energy, under Award Number DEAR0000326.

\providecommand{\noopsort}[1]{}\providecommand{\singleletter}[1]{#1}%
\end{document}